\documentstyle[prl,aps,epsf]{revtex}
\begin{document} \draft
\preprint{\today} \title{Soliton Confinement and the Excitation
Spectrum of Spin-Peierls Antiferromagnets
\thanks{For the proceedings of the NATO ASI on ``Dynamical properties of
unconventional magnetic systems'', Geilo, Norway, April 2-12, 1997.}}
\author{Ian Affleck} \address{Department of Physics and Canadian Institute
for Advanced Research, \\ University of British Columbia, Vancouver, BC,
V6T 1Z1, Canada} \maketitle \begin{abstract}  The excitation spectrum of
spin-Peierls antiferromagnets is discussed taking into acount phonon
dynamics but treating inter-chain elastic couplings in mean field theory. 
This gives a ladder of soliton-anti-soliton boundstates, with no soliton
continuum, until soliton deconfinement takes place at a transition into a
non-dimerized phase.

  \end{abstract} 

Much of the theoretical work on spin-Peierls systems largely ignores
phonon dynamics, regarding the lattice distortion as being static,
thus producing an alternating Heisenberg exchange coupling.  In the
approach of Cross and Fisher \cite{Cross}  the resulting susceptibility
of the spin system (its response to a lattice distortion) is then fed
into an RPA calculation of the phonon Green's function.  However, the
magnetic excitations, at least in the dimerized phase, are regarded as
containing no phonon component.  On the other hand, some
theoretical work on the magetic field or impurity induced undimerized
phase {\it has} considered solitons which are at the same time
magnetic excitations (they have spin 1/2) and involve lattice
dynamics (the lattice distortion switches between the two phases at the
location of the soliton).  Khomskii et al.\cite{Khomskii} have
developed a simple appealing picture of the spin-Peierls transition
(as a function of field, impurity concentration or temperature)  based
on soliton unbinding.  Their approach differs fundamentally from that
of Cross and Fisher in its treatment of phonon dynamics.  Khomskii et
al. attempt to treat the one-dimensional phonon dynamics accurately,
while approximating the {\it interchain} elastic couplings by mean
field theory.  This approach probably works best when the transverse
phonon dispersion energy is small compared to the magnetic energy
gap.  Such an approach leads to an effective one-dimensional model
containing both spontaneous dimerization due to one-dimensional
phonons and explicit dimerization produced by the mean field of the
neighbouring chains.  This is to be contrasted with the other approach
which only contains explicit dimerization.  In the approach of
Khomskii et al. the fundamental excitations of the system are
solitons.  In the spin-Peierls ordered phase the solitons ($s$)  and
anti-solitons ($\bar s$) are bound together in pairs by a linear
potential arising from the explicit dimerization potential of the
neighbouring chains.  A sufficient temperature, field or impurity
concentration drives this self-consistently determined dimerizing
field to zero, eliminating the linear potential between solitons,
allowing free solitons to propogate.  From this perspective the
spin-Peierls transition corresponds essentially to soliton deconfinement
rather than phonon softening.  

In this paper we wish to extend
this approach to a more quantum mechanical treatment of the
excitations in the spin-Peierls ordered phase.  The $s\bar s$ system,
with its linear potential is quantized, leading to a ladder of
boundstates which can have spin 0 or 1.  The  s=1 boundstates
correspond to  magnons.  As the system becomes more one-dimensional,
the linear potential gets weaker and the number of stable boundstates
increases, diverging in the one-dimensional limit.  The s=0
boundstates could also be interpreted as boundstates of two magnons, 
but in the highly one-dimensional case they are better interpreted as
weakly bound singlet $s\bar s$ pairs with energies given by twice the
energy of the soliton plus a (positive) ``binding energy'' associated
with the linear potential.  All boundstates are both magnetic and
elastic in character; the s=0 boundstates are not neccessarily
distinct from optical phonons. 

 Within this approach, a soliton continuum of
excitations does not occur in the ordered phase, contrary to the
claims in \cite{Ain}. Instead the continuum is quantized into a ladder
of boundstates.  A {\it magnon} continuum can occur, beginning at
precisely twice the magnon gap.  All excited $s\bar s$ boundstates
must lie {\it below} this continuum (in the appropriate spin channel)
in order to be stable.  The occurance of $s\bar s$ boundstates in
Heisenberg models with competing spontaneous and explicit dimerization
was mentioned by Haldane \cite{Haldane}.  The fact that a soliton
continuum cannot occur in spin-Peierls systems due to soliton
confinement was pointed out recently by Uhrig and Schulz \cite{Uhrig}.

The occurence of a ladder of $s\bar s$ boundstates
is generic to quasi-one-dimensional systems with broken discrete
symmetries, at least within this type of mean field treatment.  A
very similar approach was taken by Shiba \cite{Shiba} to  Ising
antiferromagnets who referred to the excitations as a ``Zeeman
ladder''.  These excitations have apparently been observed in
CsCoCl$_3$  \cite{Goff}.  Such ladders of boundstates also occur in
confining (1+1) dimensional quantum field theories such as quantum
electrodynamics (Q.E.D.) \cite{Coleman}, the CP$^n$ model \cite
{Witten} and a generalized ``two-harmonic'' version of the sine-Gordon
model discussed below.  Our approach to the spin-Peierls problem was
partly inspired by the work of Coleman \cite{Coleman} on weakly coupled
Q.E.D.  

We begin by considering a simple
one-dimensional s=1/2 antiferromagnetic model without phonons:
\begin{equation} H=\sum_i[J(1+\delta (-1)^i)\vec S_i\cdot \vec
S_{i+1}+J_2\vec S_i\cdot \vec S_{i+1}].\label{H1}\end{equation}
We first discuss the case $\delta =0$.  For $J_2<J_{2c}\approx .24 J$
the model is in a non-dimerized gapless phase.  On the other hand,
for larger $J_2$ the groundstate is spontaneously dimerized and there
is a gap \cite{Haldane,Chitra}.  Thus, in principle, spontaneous
dimerization could occur even without phonons.  There is some evidence
that the value of $J_2$ in CuGeO$_3$ may be close to
$J_{2c}$\cite{Castilla} but clearly phonons  also play an important
role in the spin-Peierls transition.  We include $J_2$ here
for a different reason.  In order to capture the essential physics of
spontaneous dimerization without explicitly including phonons, it is
neccessary to choose $J_2>J_{2c}$.   

In the limit where $J_2$ is only
slightly larger than the critical value the low energy excitation
spectrum can be determined by bosonization\cite{Affleck1}.  This gives
the sine-Gordon model with Hamiltonian density: \begin{equation} {\cal
H}=v\left[ {\beta^2\over 8}\Pi^2+{2\over \beta^2}\left({d\phi \over
dx}\right)^2 +g\cos 2\phi \right],\label{HSG}\end{equation} with 
$8\pi -\beta^2$ and $g$  $\propto (J_{2}/J_{2c}-1)$. 
Since this interaction has renormalization group scaling dimension
$\beta^2/4\pi$ it is marginally relevant for $J_2>J_{2c}$ leading to a
soliton gap $\Delta_s \propto \exp [-\hbox{const}/(J_2-J_{2c})]$. It
follows from the bosonization procedure that $\phi$ is to be
interpreted as an angular variable and that the discrete symmetry
$\phi \to \phi + \pi$ corresponds to translation by one site. 
Classically there are two groundstates at $\phi=\pm \pi /2$,
corresponding to spontaneously broken translational symmetry. The $s$
and $\bar s$ interpolate between these groundstates and have spin
$S^z=\pm 1/2$ depending on whether $\phi$ rotates clockwise or
counterclockwise with increasing $x$. These groundstates may be
simplistically pictured as consisting of nearest neighbour dimers in
one of the two possible patterns and the $s$ or $\bar s$ as being a
single unpaired spin separating the two different dimer patterns, as
shown in Figure 1. The
(presumably) exact results on the sine-Gordon model indicate that
there are no other excitations besides the solitons and anti-solitons
(and of course their multi-particle states) for this range of
$\beta$ \cite{Dashen}.  In particular, unlike in a
perturbative treatment in $\beta$, there are no approximately harmonic
excitations in addition to the topological ones. This is presumably a
rather special feature of this particular model.  The soliton width
and spin correlation length  obey $\xi_s \propto v/\Delta_s$,
diverging at $J_{2c}$.  As we increase $J_2$ further this soliton
width decreases.  At the special point $J_2=J/2$ the exact
groundstate is given by nearest neighbour dimers.  In this entire
region the translational symmetry is spontaneously broken so it
follows from general principles that the excitation spectrum must
contain solitons.  

\begin{figure}
\epsfxsize=8.1cm \centerline{\epsffile{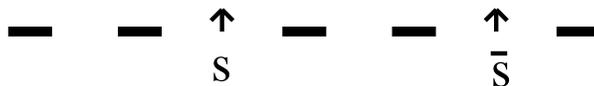}}
\caption{An $s\bar s$ pair. The solid line represents a dimer singlet.} 
\label{fig:s}
\end{figure}

Next we consider the model of Eq. (\ref{H1}) with a small non-zero
$\delta$ orresponding to a mean field from the neighbouring chains
preferring one of the two dimerization patterns.  The staggered term
leads to an additional sine-Gordon interaction: \begin{equation}
{\cal H}=v\left[ {\beta^2\over 8}\Pi^2+{2\over \beta^2}\left({d\phi \over
dx}\right)^2 +g\cos 2\phi +c\delta
\sin \phi \right] ,\label{HSG2}\end{equation} 
where $c$ is a constant of $O(1)$.
 The spectrum is fundamentally different depending on the
value of $J_2$.  For $J_2<J_{2c}$ we may ignore the
marginally irrelevant $\cos 2\phi$ interaction in
Eq. (\ref{HSG2}) arising from the uniform terms in Eq. (\ref{H1}). 
On the other hand, the $\sin \phi$ interaction  has
dimension 1/2 and thus produces a gap $\propto \delta^{2/3}$ (up to
logarithmic corrections).  The exact spectrum consists of a triplet and
a higher energy singlet with
$\Delta_1/\Delta_3=\sqrt{3}$ \cite{Dashen,Haldane}.

For $J_2>J_{2c}$, 
there is  a competition between spontaneous and explicit
dimerization, represented by the $g\cos 2\phi$ and $c\delta \sin \phi$
terms respectively in the sine-Gordon Hamiltonian.  If $\delta $ is
very small then, while the true groundstate has $\phi = -\pi /2$ the
other state at $\phi = \pi /2$ has only slightly higher energy. 
Consider a $s\bar s$ configuration
where the true groundstate ($-\pi /2$) occurs at $x\to \pm \infty$ but
the unstable groundstate occurs in between the $s\bar s$ pair,
separated by a distance $x$.  Defining the $s$ as the kink with
 $\phi = -\pi /2$ at $x\to -\infty$ and the $\bar s$ as the kink
with $\phi = \pi /2$ at $x\to -\infty$, we see that the $s$ must
always be to the left of the $\bar s$. If $x$ is much greater than the
soliton width $\xi_s$ then the classical energy of such a
configuration is simply $2\Delta_s+c\delta x$.  This is true
regardless of the spins of the two solitons (i.e. whether $\phi$ winds
clockwise or counter-clockwise).  This linearly confining potential is
crucial to the physics of spin-Peierls systems or indeed of any system
with a competition between spontaneous and explicit breaking of a
discrete symmetry.  

We now wish to treat this model quantum mechanically, in
the small $\delta$ limit.  As far as we know, this ``2-harmonic
sine-Gordon model'' model is not integrable.  However, for small $\delta$
we may follow Coleman's treatment of weakly coupled (1+1)-dimensional
Q.E.D.\cite{Coleman}.  Consider a single $s\bar s$ pair.  We expect it to
give a spectrum of boundstates, due to the linear potential.  Thus we
write an effective Hamiltonian for the center of mass co-ordinate, $x$, of
the pair: \begin{equation} H=-{1\over M_s}{d^2\over dx^2}+c\delta
x,\label{HNR}\end{equation} 
where $M_s\ \approx \Delta_s/v^2$ is the soliton
mass. We  restrict $x\geq 0$ and impose a vanishing boundary condition
on the wavefunction at $x=0$. The most important point about the
eigenstates of this Hamiltonian is that they consist entirely of
boundstates; there is no continuum.  This follows because the
potential keeps on increasing for all $x$ so the soliton and
anti-soliton never escape to infinity.  A good idea of the nature of
the spectrum can be obtained from the WKB approximation.  The
$n^{\hbox{th}}$ eigenvalue is given by:
\begin{equation} \int_0^{x_0}dx\sqrt{M_s(E^B_n-c\delta x)}=n\pi
,\end{equation} where $x_0$ is the classical turning point,
$x_0\equiv E^B_n/c\delta $.  Thus the number of states with energy less
than $E^B$ is:
\begin{equation} N(E^B)\approx {2\sqrt{M_s}(E^B)^{3/2}\over c\delta
3\pi}.\end{equation}
We see that the density of states per unit energy diverges as $\delta
\to 0$.  This is another peculiarity of a linear potential.  As the
strength of the potential goes to 0 the free particle limit is
obtained by the boundstates becoming more and more dense until they
fill in the $s\bar s$ continuum.  We have attached the superscript on
$E^B$ to remind the reader that this is the binding energy of the
$s\bar s$ pair; the total energy consists of this binding energy
together with the $s\bar s$ rest mass energy:
\begin{equation} E_n = 2\Delta_s+E^B_n.\end{equation}

Actually, the results of the previous paragraph are only valid for
low energy boundstates with $E^B<<\Delta_s$ where the
non-relativistic approximation to the soliton dispersion relation may
be used.  Coleman extended the validity of this result by using the
relativistic version of the WKB approximation.  This is certainly
valid for Q.E.D.  We also expect it to be valid for our
antiferromagnetic chain in the limit where $c\delta <<\Delta_s<<J$, 
in which the theory is approximately Lorentz invariant up to the energy
scale $\Delta_s$.  Thus the free soliton energy may be written:
\begin{equation} E(p)=\sqrt{\Delta_s^2+v^2p^2}.\end{equation}  The
WKB condition now becomes:
\begin{equation} \int dpdx\theta (E_n-2E(p)-c\delta
x) = {1\over c\delta}\int dp[E_n-2E(p)]\theta
[E_n-2E(p)]=2\pi n.\label{WKBR}\end{equation}  Here $E_n$ is the full
relativistic energy of the $s\bar s$ pair including both kinetic
energy of the individual solitons and binding energy. So far this
discussion ignores many body effects.  In the weak coupling limit, as
Coleman observed, these simply truncate this boundstate spectrum at
$E<4\Delta_s$.  Any $s\bar s$ boundstate of higher energy is unstable
because it can decay into a pair of boundstates. If we imagine trying
to pull an $s\bar s$ pair apart to $\infty$ it eventually becomes
energetically favourable for ``pair production'' to occur so that we
end up separating to $\infty$ two $s\bar s$ pairs.  Setting
$E=4\Delta_s$ in Eq. (\ref{WKBR}) gives the number of stable
boundstates: \begin{equation} N\approx .684
\Delta_s/ c\delta.\end{equation}  We note that this $1/\delta$
behavior is independent of the free soliton dispersion relation
(although the prefactor depends on it).  Shiba\cite{Shiba} encountered
essentially a discrete lattice version of this Schroedinger equation
in his work on Ising antiferromagnets, coming to similar conclusions
about the spectrum.

So far we have ignored the soliton spin.  The Hamiltonian of Eq.
(\ref{HNR}) is spin-independent so both the $s$ and
$\bar s$ can independently have spin up or down.   Thus each of these
boundstates corresponds to a degenerate triplet and singlet.  When the  s
and $\bar s$ are close together (on the scale $\xi_s$) their interaction
will be a good deal more complicated; in particular it will be spin
dependent.  Fortunately, for very small $\delta$ they stay far apart
even in the lowest boundstate.  Note that the $n=1$ classical turning
point is at $x_0\propto 1/\delta^{1/3}$. As we increase $\delta$ we
expect the number of boundstates to decrease and the degeneracy
beween triplet and singlet boundstates to be lifted. 
The set of stable boundstates must
always lie below the two boundstate continuum in the coresponding spin
channel. Clearly free solitons can never appear in the spectrum for
non-zero $\delta$.  Eventually, when $\delta >>(\Delta_s/J)^{3/2}$ (but
still $\delta <<1$) we expect to recover the spectrum of the
$J_2<J_{2c}$ model with one triplet and one singlet with the ratio
$\Delta_1/\Delta_3=\sqrt{3}$.  It would be interesting to study the
spectrum of the spin model numerically to test these ideas.  (The
expected behavior for $J_2<J_{2c}$ and $\delta <<1$ was recently
confirmed \cite{Augier}.)

We now  include phonons in our model, taking the Hamiltonian:
\begin{equation}H=\sum_i[(J+\alpha u_i)\vec S_i\cdot \vec
S_{i+1}+\Pi_i^2/M + (K/2)u_i^2+\delta
(-1)^iu_i].\label{HSP}\end{equation}
Here $u_i$ is the change in  separation of two
neighbouring ions from its uniform value and $\Pi_i$ is the
corresponding conjugate momentum; $M$ is the ionic mass. (We have
omitted acoustic phonons for simplicity.  They presumably are not
important for the spin-Peierls effect.)  $\delta$ represents a mean
field arising from the coupling to neighbouring chains which favors
one of the two possible lattice distortions. We may also keep a second
nearest neighbour exchange coupling; it doesn't change the discussion
qualitatively. While this Hamiltonian is considerably more complicated
than the spin-only one considered above, we expect many features of
the previous discussion to carry over.  When $\delta=0$ we expect two
degenerate groundstates with $<u_i>=\pm u_0(-1)^i$.  The excitation
spectrum will include s=1/2 solitons and anti-solitons. 
 Note that all
excitations now involve both spin and phonon degrees of freedom.  In
addition to solitons, the excitation spectrum will presumably include
other excitations corresponding to optical phonons.  In principle
there might also be integer spin magnetic quasi-particles below the
two-soliton continuum.  However, since the interaction between $s$
and $\bar s$ vanishes at long distances for $\delta =0$, no boundstates
need occur in this limit, as in the previous spin-only model. Turning on a
small $\delta$ will again confine the solitons into $s\bar s$ pairs.  In
the small $\delta$ limit we expect the number of stable boundstates
to be proportional to $1/\delta $. Presumably spin 0 $s\bar s$
boundstates can mix with optical phonons, which may reduce the
number of stable s=0 boundstates.  As $\delta$ increases the number
of boundstates will decrease.  Free solitons can never appear in the
spectrum.  In addition to the stable boundstates there will also be 
two boundstate continua in the various spin channels (and also
presumably a two phonon continuum).  Evidently all boundstates must
lie below the continuum in the corresponding spin channel in order to
be stable.  Numerical results on such a one-dimensional spin-phonon
system would be highly desirable although considerably more difficult
than for a spin-only system. 

Finally, let us consider a full three dimensional spin-Peierls
Hamiltonian.  A simple model would consist of chains with the
Hamiltonian of Eq. (\ref{HSP}) (with $\delta =0$) together with an
inter-chain phonon coupling:
\begin{equation} K'\sum_i\sum_{<\vec R, \vec R'>}u_{i\vec R}u_{i\vec
R'},\end{equation} where $\vec R$ labels the lattice points
perpendicular to the chains.  In the ordered phase $<u_{i\vec
R}>=(-1)^iu_0$.  Treating the interchain coupling in mean field
theory gives the one dimensional model of Eq. (\ref{HSP}) with
$\delta = ZK'u_0$ where $Z$ is the chain co-ordination number. 
$u_0$ should then be determined self-consistently by
solving the one-dimensional model.   This
is essentially the approach advocated by Khomskii et al.
\cite{Khomskii}.   Note that it is a rather standard
approach to various quasi-one-dimensional systems \cite{Affleck2}.
 In the disordered phase, $\delta
=0$ so solitons can exist as independent excitations.  In the ordered
phase $\delta >0$ and free solitons cannot occur due to the confining
potential. While based on a mean field treatment of inter-chain
couplings this conclusion is presumably much more general.
It essentially follows from Landau's argument
that any non-zero density of free solitons leads to the destruction of
long range order in a one dimensional system.  The theoretical discussion
in Ref. \cite{Ain} essentially considered a purely one-dimensional spin-
phonon model which does indeed contain free solitons.  The magnon was
regarded as an $s \bar s$ boundstate that could occur in this model.
  However, it is
not permissible to ignore inter-chain elastic coupling in the dimerized
phase.  This coupling at the same time stabilizes the dimerized phase up 
to a finite crticial temperature and confines the solitons.
Note that in the small
$K'$ limit, the energy scale of all excitations is set by the
one-dimensional model and therefore the magnon gap should be given by
 twice the soliton gap (assuming no boundstates in the
one-dimensional model for $\delta =0$).  Even if $K'$ is not very
small, its effects become smaller near a transition into a
non-dimerized phase, driven by temperature, field or impurity
concentration.  Thus, at least naively, one might expect the present
approach to become more valid near such a transition.  

What does this approach tell us about CuGeO$_3$? We may 
interpret the observed magnon as a spin 1 $s\bar s$ boundstate. 
Likewise, if there is a stable singlet quasi-particle, as suggested
by Raman scattering \cite{Muthukumar}, we may interpret it as an s=0
$s\bar s$ boundstate.  As remarked above, there appears to be no sharp
distinction, in general, between such an s=0 $s\bar s$ boundstate and
an optical phonon in a spin Peierls system.  In Raman scattering the
photon couples both to lattice displacements and to $\vec S_i\cdot \vec
S_{i+1}$.  Such spin 0 excitations might also be observable in
neutron scattering because they couple to lattice displacements which
are excited by neutron scattering from the ionic nuclei.  One precise
conclusion from the present approach is that there can be no soliton
continuum in the spin-Peierls phase.  The continuum in the magnetic
neutron scattering cross-section neccesarily starts at twice the gap
to the lowest spin triplet $s\bar s$ boundstate corresponding 
to a 2-magnon continuum.  This is consistent with the data presented
in \cite{Ain}.  The apparent absence of additional $s\bar
s$ boundstates in CuGeO$_3$ (besides the magnon and possibly one
singlet) can presumably be attributed to the relatively large value of
the inter-chain phonon coupling, $K'$. Likewise the failure to
observe an $s\bar s$ continuum above $T_{SP}$ can be so attributed
since then $T_{SP}$ is of the order or greater than $\Delta_s$,
smearing the continuum threshold.  It would be interesting to find
  spin-Peierls materials which were more highly one-dimensional with
respect not only to their magnetic exchange couplings but also their
elastic couplings.  In such materials additional  $s\bar s$ boundstates
should exist below $T_{SP}$ and the $s\bar s$ continuum could perhaps
be observed above $T_{SP}$.  One way in which enhanced one-dimensionality
might occur is in a system where the next nearest neighbour Heisenberg
coupling $J_2>J_{2c}\approx .24J$.  In this case the soliton gap could be
determined primarily by the magnetic exchange energies and might be large
 compared to $K'$ or other phonon energy scales.  

I would like to thank Michel A\"\i n, Collin Broholm, Roger Cowley and
Michael Fischer for helpful discussions.  This research was supported by
NSERC of Canada.

  \end{document}